\documentclass[12pt]{article} 
\usepackage{graphicx}
\usepackage{amsmath}
\usepackage{amssymb}
\allowdisplaybreaks

\usepackage{utopia}
\usepackage[utopia]{mathdesign}
 \usepackage{cite}
 \usepackage{hyperref}
\usepackage[height=8.85in,width=6.45in]{geometry}
\numberwithin{equation}{section}

\newcommand{\cG}{\mathcal{C}}
\def\Nequals#1{$\mathcal{N}{=}#1$}

\newcommand{\bZ}{\mathbb{Z}}

\def\ket#1{\left| #1 \right\rangle}

\def\vev#1{\left\langle #1 \right\rangle}

\def\U{\mathrm{U}}
\def\SU{\mathrm{SU}}

\def\SO{\mathrm{SO}}
\def\Spin{\mathrm{Spin}}

\def\SU{\mathrm{SU}}
\def\SL{\mathrm{SL}}
\def\cH{\mathcal{H}}
\let\oldmid\mid
\def\mid{\,\oldmid\,}

\def\tr{\mathop{\mathrm{tr}}\nolimits}
\def\diag{\mathop{\mathrm{diag}}}
\def\rank{\mathop{\mathrm{rank}}}

\begin{document}

\begin{titlepage}

\begin{flushright}
IPMU-14-0350\\
UT-14-47\\
\end{flushright}

\vskip 2cm

\begin{center}
{\Large \bfseries
Magnetic discrete gauge field in the confining vacua \\[1em]
and the supersymmetric index \\[3em]

}

\vskip 1.5cm
 Yuji Tachikawa
\vskip 1.0cm

\begin{tabular}{ll}
  & Department of Physics, Faculty of Science, \\
& University of Tokyo,  Bunkyo-ku, Tokyo 133-0022, Japan, and\\
  & Institute for the Physics and Mathematics of the Universe, \\
& University of Tokyo,  Kashiwa, Chiba 277-8583, Japan\\
\end{tabular}

\vskip 1cm

\textbf{Abstract}

\end{center}

\medskip
\noindent
It has recently been argued that the confining vacua of Yang-Mills theory in the far infrared can have topological degrees of freedom given by magnetic $\bZ_q$ gauge field, both in the non-supersymmetric case and in the \Nequals1 supersymmetric case.  In this short note we give another piece of evidence by computing and matching the supersymmetric index of the pure super Yang-Mills theory both in the ultraviolet and in the infrared.

\end{titlepage}



\section{Introduction and Summary}



There have been many ideas proposed to explain the mechanism of the color confinement.
  One influential idea has been  the monopole condensation  \cite{'tHooft:1977hy,'tHooft:1979uj}, that the color is confined due to the condensation of magnetically charged objects.

Let us for a moment  consider a case where a $\U(1)$ gauge symmetry is broken by a scalar field of electric charge $q$.  In the infrared, there still is an unbroken $\bZ_q$ gauge symmetry. Such a discrete gauge field is locally trivial, but has a subtle physical effect globally.  For example, in a conventional superconductor, the Cooper pairs have charge 2, and therefore there is a $\bZ_2$ gauge symmetry.

In a class of confining gauge theories, such as softly-broken \Nequals2 supersymmetric systems, the confinement proceeds in two steps\cite{'tHooft:1981ht,Seiberg:1994rs}: the gauge group is effectively broken to its maximal Abelian subgroup by the strong dynamics, which is then confined by the condensation of monopoles.  We see that, if the monopoles have charge greater than 1, there can be a \emph{magnetic $\bZ_q$ gauge symmetry} in the confining vacuum.  

The appearance of magnetic $\bZ_q$ gauge symmetry is a much more general phenomenon, independent of whether the confinement proceeds as above. For example, we argued in \cite{Aharony:2013hda} that pure (non-supersymmetric) Yang-Mills theory with gauge group $\SU(N)/\bZ_N$, with the theta angle $\theta=2\pi k$ where $k$ is an integer\footnote{Note that the instanton number of $\SU(N)/\bZ_N$ gauge fields on a nontrivial spin manifold is $1/N$ times an integer, and therefore the periodicity of the theta angle is $\theta\sim \theta+2\pi N$.}, has a magnetic $\bZ_{\gcd(N,k)}$ gauge symmetry in the infrared confining vacuum.  

A rough argument, using the monopole condensation picture, goes as follows.  Say, in the $\SU(N)$ Yang-Mills theory with $\theta=0$,  the confinement is due to the magnetic $\U(1)$ gauge field broken by a condensate of magnetic charge $1$. 
In the $\SU(N)/\bZ_N$ theory, the periodicity of the magnetic $\U(1)$ gauge field changes by a factor of $N$. Stated  differently, the magnetic $\U(1)$ is now broken by a condensate of magnetic charge $N$, thus giving a magnetic $\bZ_N$ gauge field in the infrared. 
Now, by shifting the theta angle to $\theta=2\pi k$, there is a discrete version of the Witten effect \cite{Witten:1979ey}, and the condensate has the magnetic charge $N$ and the electric charge $k$.  Therefore, what remains in the infrared is a  $\bZ_{\gcd(N,k)}$ gauge field. 
The same conclusion can also be reached without using the monopole condensation picture \cite{Aharony:2013hda}.

It can be said that the confining vacuum of the Yang-Mills theory is a version of \emph{symmetry protected topological phase}, if the reader allows the author to use a more fashionable terminology used these days. This point of view is further studied in e.g.~\cite{Kapustin:2013uxa,Kapustin:2014gua,GreatArticleToAppear}.

The aim of this short note is to provide another piece of evidence to the existence of this magnetic $\bZ_q$ gauge field, by considering the Witten index of \Nequals1 supersymmetric pure  Yang-Mills theory with various gauge group $G$. We start by studying the simplest cases when $G=\SU(2)$ and $G=\SO(3)$ in Sec.~\ref{sec:simplest}.  
In Sec.~\ref{sec:sun} and in Sec.~\ref{sec:son}, we analyze the cases $G=\SU(N)/\bZ_N$ and $G=\SO(N)$, respectively. Finally we analyze pure super Yang-Mills theory with arbitrary connected gauge groups in Sec.~\ref{sec:general}.  

It should be remarked at this point that all of the difficult gauge-theoretic computations that are required for the analysis of this note have already been performed in \cite{Borel:1999bx,Witten:2000nv}, and what will be presented below is just a translation of the result (3.31) in \cite{Witten:2000nv} in the language of \cite{Aharony:2013hda}. Therefore, there is nothing new in this note, except for a possibly new viewpoint that emphasizes the magnetic $\bZ_q$ gauge field in the confining vacuum.

\section{With $G=\SU(2)$ and $G=\SO(3)$ }\label{sec:simplest}

\paragraph{When $G=\SU(N)$.} In a classic paper \cite{Witten:1982df}, Witten considered pure \Nequals{1} super Yang-Mills theory with $G=\SU(N)$ by making the spatial slice to be $T^3$ of size $L$. Let us first briefly recall the analysis performed there. 

The spatial slice is periodically identified, $x_i\sim x_i+L$ for $i=1,2,3$.
We denote by $Z(L)$  the Witten index of the system \begin{equation}
Z(L)=\tr _\cH (-1)^F e^{-\beta H}
\end{equation} where $\cH$ is the Hilbert space, $F$ the fermion number, and $H$ is the Hamiltonian. Using the by-now standard argument, we know $Z(L)$ is independent of $L$.   

When $L$ is very, very big, we can compute $Z(L)$ using the structure of the infrared vacua. 
The system has an $\bZ_{2N}$ R-symmetry which acts on the gaugino $\lambda$  by a multiplication by an $2N$-th root of unity.  The gaugino condensate $\vev{\tr \lambda\lambda}$ breaks it to $\bZ_{2}$, which is the $360^\circ$ rotation of the space. Therefore, there are $N$ vacua related by the action of the R-symmetry, with the gaugino condensate \begin{equation}
\vev{\tr\lambda\lambda}=\Lambda^3, \quad \omega \Lambda^3, \quad \ldots,\quad \omega^{N-1} \Lambda^3
\end{equation} where $\omega=e^{2\pi i/N}$  and $\Lambda$ is the gauge theory dynamical scale. 

By putting the system on a large $T^3$, these $N$ vacua give $N$ zero energy states. They  all have   the same value of $(-1)^F$, since they are related by the action of the R-symmetry. We thus find \begin{equation}
| Z_{\SU(N)}(L) |  =  N \qquad (L\Lambda \gg 1).\label{foo}
\end{equation}

When $L$ is very, very small, the system is weakly-coupled, and the index $Z(L)$ can be computed reliably using semi-classical methods. To have zero energy states, the  holonomies $g_{1,2,3}\in \SU(N)$ around the three nontrivial paths $x_i \sim x_i +L$ of $T^3$ need to commute. 
 They can be simultaneously conjugated into the Cartan torus $T\subset \SU(N)$. 
 
 The system is then effectively described by a supersymmetric quantum mechanics with the following structure:  The  bosonic degrees of freedom are  $g_{1,2,3}\in T$, the fermionic degrees of freedom are $\lambda_{1,2} \in \mathfrak{t}$ where $\mathfrak{t}$ is the Lie algebra of $T$, and we need to impose the invariance under the Weyl symmetry $S_N$.
The zero-energy states are then given by \begin{equation}
\ket 0, \quad (\tr \lambda_1\lambda_2)\ket 0,\quad (\tr \lambda_1\lambda_2)^2\ket 0, \quad\ldots,\quad (\tr \lambda_1\lambda_2)^{N-1}\ket 0
\end{equation} with a suitably chosen state $\ket 0$;
note that $\tr(\lambda_1\lambda_2)^N=0$ because $\rank T=N-1$.

In the end  we find \begin{equation}
| Z_{\SU(N)}(L) |  =  1+\rank T=N, \qquad (L\Lambda \ll 1).
\end{equation} This is consistent with what we found in the infrared, \eqref{foo}. 

\paragraph{When $G=\SO(3)$.}  Let us now consider what changes when we consider $G=\SU(N)/\bZ_N$. The case with general $N$ will be considered momentarily; let us first study the simplest case $N=2$. 

We begin by considering when the system size $L$ is very very small.  As before, we need to analyze the supersymmetric quantum mechanics based on three commuting holonomies $g_{1,2,3}\in \SO(3)$. We still have the component when $g_{1,2,3}\in T\in \SO(3)$ where $T$ is the Cartan torus. This still gives $N=2$ states as before. 

But this is not all. We can take, for example, three matrices \begin{equation}
g_1=\diag(+1,-1,-1),\quad
g_2=\diag(-1,+1,-1),\quad
g_3=\diag(-1,-1,+1)
\end{equation}  that mutually commute but cannot be in the same Cartan torus. In fact this is isolated and its gauge equivalence class cannot be continuously deformed.  This gives one zero-energy state. 

Lifting from $\SO(3)$ to $\SU(2)$, we find that the holonomies $g_{1,2,3}$ lift to Pauli matrices $\sigma_{1,2,3}$.  Note that $g_1g_2=g_2g_1$ but $\sigma_1\sigma_2=-\sigma_2\sigma_1$.
This means that the Stiefel-Whitney class\footnote{This is the $w_2$ of the gauge bundle. In this note we only consider tori with trivial spin structure.} $w_2$ of the $\SO(3)$ bundle, evaluated on the face $C_{12}$ of the $T^3$, gives $-1$. 
Here and in the following,  $C_{ij}$ is the $T^2$ formed by the edges in the $i$-th and the $j$-th directions of $T^3$.  
We can similarly compute $w_2(C_{23})$ and $w_2(C_{31})$; we have  $(w_2(C_{23}),w_2(C_{31}),w_2(C_{12}))=(-1,-1,-1)$.

In general, the possible choices of $w_2$ are $(\pm1,\pm1,\pm1)$. The commuting triples in the class $(+1,+1,+1)$ are the ones that can be simultaneously conjugated to the Cartan torus $T\subset \SO(3)$ discussed above, and they give 2 states.   For each of the other seven choices of $w_2$, there is one  isolated commuting triple, that gives one zero-energy state.\footnote{\label{UVfootnote}The fermion number $(-1)^{F_1}$ of these seven states  is the same as the fermion number $(-1)^{F_0}$ of the two states we found earlier. To see this, let us consider the partition function on a small $T^4$  with fixed $w_2$.
When $w_2$ is trivial along the spatial $T^3$, the $T^4$ partition function has the phase $(-1)^{F_0}$, independent of $w_2$ along the temporal-spatial directions. 
When $w_2$ is nontrivial along the spatial $T^3$ but trivial along the temporal-spatial direction, the partition function has the phase $(-1)^{F_1}$. These two configurations can be mapped to each other by exchanging the time and the space directions. 
Therefore, we should have $(-1)^{F_0}=(-1)^{F_1}$.}
In total, we find \begin{equation}
|Z_{\SO(3)}(L)|=2+7=9 \qquad(L\Lambda \ll 1).\label{bar}
\end{equation} 

Therefore, we should find the same when $L$ is very, very big. But how? There are still two vacua, with $\vev{\tr\lambda\lambda}=\pm \Lambda^3$.  But one vacuum has magnetic $\bZ_2$ gauge symmetry while the other does not \cite{Aharony:2013hda}. 
More precisely, the theory has a line operator with nontrivial $\bZ_2$ charge, coming from the 't Hooft line operator  in the ultraviolet. In the first vacuum it has a perimeter law, and in the second vacuum it has an area law.

Thus, on a very big $T^3$, the first vacuum gives $2^3$ states due to the choice of the holonomies on $T^3$, and the second vacuum gives just 1. In total\footnote{\label{IRfootnote}Again, all the states have the same $(-1)^F$. Note that the $\bZ_2$ gauge theory on $T^3$ has a global symmetry $\cG:=H^1(T^3,\bZ_2)$, given by tensoring the gauge bundle by another $\bZ_2$ bundle. The charge under $\cG$ is $\cG^\vee=H^2(T^3,\bZ_2)$. 
Now, the $2^3$ states coming from the first vacuum are permuted by $\cG$; let us say they have $(-1)^F=(-1)^{F_a}$.  The additional state from the second vacuum is invariant under $\cG$, with $(-1)^F=(-1)^{F_b}$.  Stated differently, there are one state with $(-1)^F=(-1)^{F_a}$ for each charge in $\cG^\vee$, and another state with $(-1)^F=(-1)^{F_b}$ with zero charge in $\cG^\vee$.
Now, the two states with zero charge in $\cG^\vee$ are the same two states in the $\SU(2)$ theory, and therefore have the same $(-1)^F$. Therefore we see that $(-1)^{F_a}=(-1)^{F_b}$.}, we find \begin{equation}
|Z_{\SO(3)}(L)|=2^3+1=9 \qquad(L\Lambda \gg 1).
\end{equation} 
This is again consistent with the computation in the opposite regime \eqref{bar}.

\section{With $G=\SU(N)/\bZ_N$}\label{sec:sun}
Now let us move on to the case $G=\SU(N)/\bZ_N$. The index of the system in a large $T^3$ can be found easily. 
There are $N$ vacua in the infinite volume limit, and as discussed in \cite{Aharony:2013hda} and recalled in the Introduction, the $k$-th vacuum has magnetic $\bZ_{\gcd(N,k)}$ symmetry. Each vacuum with $\bZ_q$ symmetry gives rise to $q^3$ zero energy states in a large $T^3$. Therefore the Witten index is\footnote{That all the states have the same value of $(-1)^F$ can be seen exactly as explained in footnote \ref{IRfootnote}.} \begin{equation}
|Z_{\SU(N)/\bZ_N}(L)| = \sum_{k=1}^N \bigl(\gcd(N,k)\bigr)^3=\sum_{m\mid N} (N/m)^3 \varphi(m), \qquad (L\Lambda \gg 1)\label{bok}
\end{equation} where $m|N$ denotes that $N$ is divisible by $m$, and $\varphi(m)$ is  Euler's totient function, i.e.~the number of positive integers less than $m$ and relatively prime with $m$.

To perform the computation in the opposite regime, we need to understand the moduli space of commuting triples of $\SU(N)/\bZ_N$. 
First, the topological class of $\SU(N)/\bZ_N$ bundles on $T^3$ is labeled by its generalized Stiefel-Whitney class $w_2$. In the case of a flat bundle on $T^3$, we first take three holonomies $g_{1,2,3}\in \SU(N)/\bZ_N$ along three edges $X_{1,2,3}$ of $T^3$. We then lift each element to $\SU(N)$ and call them $h_{1,2,3}$.  Then they should commute up to the center of $\SU(N)$, i.e.~\begin{equation}
h_i h_j = m_{ij} h_j h_i \label{baz}
\end{equation}
 where $m_{ij}$ is an $N$-th root of unity.  This $m_{ij}$ is  $w_2$ evaluated on the face $C_{ij}$. 
 The topological class of the bundle is then given by \begin{equation}
(w_2(C_{23}),w_2(C_{31}),w_2(C_{12}))=(m_{23},m_{31},m_{12})\in \bZ_N^3.
\end{equation}

For example, when $(m_{23},m_{31},m_{12})=(e^{2\pi i/N},1,1)$,  it is known that any $h_{1,2,3}\in \SU(N)$ that satisfy \eqref{baz} can be conjugated to  \begin{align}
h_1^{(N)}&=e^{k 2\pi i/N},\\
h_2^{(N)}&=\diag(e^{2\pi i/N},e^{2\cdot 2\pi i /N}, e^{3\cdot 2\pi i /N},\ldots,e^{N 2\pi i/N}), \\
h_3^{(N)}&=\begin{pmatrix}
0 & 1 & 0 & \cdots & 0 \\
0 & 0 & 1 & \cdots  & 0 \\
\vdots & & \ddots & \ddots & \vdots\\
0&\cdots&&0&1\\
1&0&\cdots&0&0
\end{pmatrix}
\end{align} for some integer $k$. The corresponding elements $g_{1,2,3}^{(N)}$ do not depend on $k$.
Therefore, there is just one zero energy state with $(m_{23},m_{31},m_{12})=(e^{2\pi i/N},1,1)$.

In general, by an $\SL(3,\bZ)$ transformation, we can always arrange $(m_{23},m_{31},m_{12})=(e^{ l \cdot 2\pi i /N},1,1)$ for some $l\mid N$, and the commuting holonomies $h_{1,2,3}$ can be put to the standard form \begin{equation}
h_a =  h_a^{(N/l)} \otimes s_a, \qquad s_a \in T_{l}\subset \SU(l) 
\end{equation} where $T_{l}$ is the Cartan torus of $\SU(l)$. Again, $h_1^{(N/l)}$ has $N/l$ choices, but they all project down to the same element in $\SU(N)/\bZ_N$.
Quantizing the supersymmetric quantum mechanics based on $s_{1,2,3}$, we get $1+\rank T_l=l$ states. 

We now need to count the number of triples $(m_{23},m_{31},m_{12})$ such that they can be mapped to $(e^{2\pi i l/N},1,1)$. Equivalently, we need to count the number of  triples $(x,y,z)$ of integers mod $N$  such that $\gcd(x,y,z)=l$. This is given by \begin{equation}
\sum_{k\mid  ({N}/{l})} ({N}/{kl})^3 \mu(k)
\end{equation} where $\mu(k)$  is the M\"obius function.  Then we find that,  when $L\Lambda \ll 1$, the index is\footnote{We can follow the same argument as in footnote \ref{UVfootnote} to see that these states have the same $(-1)^F$. } \begin{align}
|Z_{\SU(N)/\bZ_N}(L)|  = \sum_{l\mid N} l \sum_{k\mid  ({N}/{l})} ({N}/{kl})^3 \mu(k)
= \sum_{l\mid m\mid N} ({N}/{m})^3 l \mu({m}/{l}) = \sum_{m\mid N} (N/m)^3 \varphi(m)
\end{align} where the M\"obius inversion formula $\varphi(m)=\sum_{l\mid m} l \mu(m/l)$ was used.  This is equal to the result above \eqref{bok} of the computation in the infrared.

\section{With $G=\SO(N)$}\label{sec:son}

Let us now consider the case $G=\SO(N)$, $N\ge 7$. But let us first recall the situation when $G=\Spin(N)$, first studied in the Appendix I of \cite{Witten:1997bs}. 

The dual Coxeter number is $N-2$, and therefore, there are $N-2$ vacua in the far infrared, distinguished by the gaugino condensate \begin{equation}
\vev{\tr\lambda\lambda}=\Lambda^3, \quad \omega \Lambda^3,\quad \ldots,\quad \omega^{N-3} \Lambda^3
\end{equation} where $\omega=\exp(2\pi i/(N-2))$. Therefore when the size $L$ of $T^3$ is very big, we find \begin{equation}
|Z_{\Spin(N)}(L)|=N-2, \qquad (L\Lambda \gg 1).
\end{equation}

The commuting holonomies $(g_1,g_2,g_3)$ can be put into either of the following standard forms: \begin{equation}
g_a\in T\subset \Spin(N)
\end{equation} where $T$ is the Cartan torus of $\Spin(N)$, or \begin{equation}
g_a = g_a^{(7)} s_a
\end{equation} where $g_{1,2,3}^{(7)}$ is a lift to $\Spin(7)$ of the following $\SO(7)$ matrices \begin{equation}
\begin{array}{l}
\diag(+1,+1,+1,-1,-1,-1,-1), \\
\diag(+1,-1,-1,+1,+1,-1,-1), \\
\diag(-1,+1,-1,+1,-1,+1,-1),
\end{array} 
\end{equation}and $s_a \in T'$ where $T'$ is the Cartan torus of $\Spin(N-7)\subset \Spin(N)$ commuting with  $g_{1,2,3}^{(7)}$. 

The former component gives $1+\rank T$  zero-energy states,
and the latter component gives $1+\rank T'$ zero-energy states. In total, we find \begin{equation}
|Z_{\Spin(N)}(L)|=(\lfloor \frac{N}2 \rfloor+1)+(\lfloor \frac{N-7}2 \rfloor+1)=N-2, \qquad (L\Lambda \ll 1).
\end{equation}

Now, we move on to the case $G=\SO(N)$. In this case, there are two choices of the discrete theta angle, so there are two theories $\SO(N)_\pm$, see \cite{Aharony:2013hda}. As argued there,
in the $\SO(N)_+$ theory all vacua have $\bZ_2$ gauge symmetry for $\SO(N)_+$,
but in the $\SO(N)_-$ theory all vacua have just $\bZ_1$ gauge symmetry. 
Therefore, in the infrared, we find \begin{equation}
|Z_{\SO(N)_+} |= 8(N-2), \quad (L\Lambda \gg 1)
\end{equation} and 
\begin{equation}
|Z_{\SO(N)_-} |= (N-2), \quad (L\Lambda \gg 1).
\end{equation} 

Let us confirm this result in a computation in the ultraviolet, $L\Lambda\ll 1$.  
The topological type of the bundle is given by the Stiefel-Whitney class 
evaluated on the faces, $(m_{23},m_{31},m_{12})\in\{\pm1\}^3$. 

When $(m_{23},m_{31},m_{12})=(+1,+1,+1)$, all the commuting holonomies are obtained by projecting the $\Spin(N)$ commuting holonomies down to $\SO(N)$. Then, these give $(1+\rank T)+(1+\rank T')=N-2$ zero-energy states as before. 

For seven other choices $(m_{23},m_{31},m_{12})\neq (+1,+1,+1)$, we can always apply $\SL(3,\bZ)$ to have $(m_{23},m_{31},m_{12})=(-1,+1,+1)$. 
In \cite{Borel:1999bx} it was proved that  the commuting holonomies are either of the form 
 \begin{equation}
g_a = g_a^{(3)} s_a
\end{equation} where $g_{1,2,3}^{(7)}$ is the following $\SO(3)$ matrices \begin{equation}
\begin{array}{l}
\diag(+1,+1,+1),\quad
\diag(-1,-1,+1), \quad
\diag(-1,+1,-1), 
\end{array} 
\end{equation}and $s_a \in T''$ where $T''$ is the Cartan torus of $\SO(N-3)\subset \SO(N)$ commuting with  $g_{1,2,3}^{(3)}$, or of the form 
 \begin{equation}
g_a = g_a^{(4)} s_a
\end{equation} where $g_{1,2,3}^{(4)}$ is the following $\SO(4)$ matrices \begin{equation}
\begin{array}{l}
\diag(-1,-1,-1,-1),\quad
\diag(-1,-1,+1,+1), \quad
\diag(-1,+1,-1,+1),
\end{array} 
\end{equation}and $s_a \in T'''$ where $T'''$ is the Cartan torus of $\SO(N-4)\subset \SO(N)$ commuting with  $g_{1,2,3}^{(4)}$.

Quantization of the zero modes then give \begin{equation}
(1+\rank T'')+(1+\rank T''')=N-2
\end{equation} states for each of the seven choices $(m_{23},m_{31},m_{12})\neq (+1,+1,+1)$.
In the $\SO(N)_+$ theory they are all kept, but in the $\SO(N)_-$ theory, they have a nontrivial induced discrete electric charge $e=(m_{23},m_{31},m_{12})$ due to the non-zero theta angle. This causes these states to be projected out. 

In total, we find \begin{equation}
|Z_{\SO(N)_+} |= 8(N-2), \quad (L\Lambda \ll 1)
\end{equation} and 
\begin{equation}
|Z_{\SO(N)_-} |= (N-2), \quad (L\Lambda \ll 1)
\end{equation}  in the ultraviolet computation, agreeing with the infrared computations. 

We can similarly perform the check for $\SO(N)/\bZ_2$ or $\Spin(4N)/\bZ_2$ that is not $\SO(4N)$; the explicit descriptions of almost commuting triples in \cite{Henningson:2007dq,Henningson:2007qr} are quite useful in this regard. Instead of describing this, let us move on to a general analysis.

\section{With general connected gauge groups}\label{sec:general}

In fact we can give a uniform argument that the computations of the Witten index in the ultraviolet and in the infrared  always agree, given the facts derived in \cite{Borel:1999bx} and \cite{Witten:2000nv}, once the basic properties of the discrete theta angle and the magnetic gauge fields given in \cite{Aharony:2013hda} are taken into account.

\paragraph{Setup.}
Let us quickly recall the concepts of the discrete theta angle $\theta_\text{disc}$ and the spectral flow $\Delta$. For more details, see \cite{Aharony:2013hda} and \cite{Witten:2000nv}.
Let the gauge group be $G/K$ where $G$ is connected and simply connected and $K$ is a subgroup of the center $C$ of $G$.  The elements of $C$ label the discrete magnetic charge, while the set of irreducible representations $C^\vee$ of $C$ label the discrete electric charge.\footnote{As abstract groups $C$ and $C^\vee$ are the same, but it is useful for the author to distinguish them to make sure  that we only perform mathematically natural operations. If it confuses the reader s/he can ignore $^\vee$'s in the notation.}
The magnetic and the electric line operators in a theory with gauge algebra $\mathfrak{g}$  can then be labeled by $C\times C^\vee$.  There is a natural  pairing $\vev{\cdot,\cdot}: C\times C^\vee\to \U(1)$, so that two charges can coexist if and only if the pairing is trivial, $1\in \U(1)$.

When the gauge group is $G/K$, the allowed magnetic charges are in $K\subset C$. 
When the discrete theta angle is zero,
the magnetic charge $m$ and the electric charge $e$ of an allowed line operator in the system are of the form $(m,e)\in K\times L^\vee\subset C\times C^\vee$, where the subgroup $L^\vee\subset C^\vee$ is such that $m\in K$ and $e\in L^\vee$ satisfy $\vev{m,e}=1 \in \U(1)$ under the Dirac quantization pairing. 
Phrased differently, we have $e = 0 \in C^\vee/L^\vee = K$.

The discrete theta angle $\theta_\text{disc}$ is a $K$-linear map from $K$ to $K^\vee$, and changes the conditions on the allowed charges in $(m,e)\in C\times C^\vee$ to be \begin{equation}
m\in K, \qquad e=\theta_\text{disc} m \in C^\vee/L^\vee=K^\vee.
\end{equation}
In other words, $\theta_\text{disc}$ measures the induced electric charge  a magnetic source has.

When $K=\bZ_q$, the discrete theta angle is therefore an integer modulo $q$.  For $G=\Spin(4n)$ and $C=K=\bZ_2\times \bZ_2$, the discrete theta angle is a $2\times 2$ matrix of mod-2 integers.

There is also the continuous theta angle $\theta_\text{cont}$, which might be more familiar. For a given $G$, there is a fixed linear map  $\Delta$ from $C$ to $C^\vee$ such that the continuous change $\theta_\text{cont}\to\theta_\text{cont}+2\pi$ is equivalent to $\theta_\text{disc}\to \theta_\text{disc}+\Delta$. Here, $\Delta$ is regarded as a map from the subset $K\subset C$ to the quotient $C^\vee/L^\vee=K^\vee$. 
This $\Delta$ is called the spectral flow and computed for all $G$ in \cite{Witten:2000nv}.

\paragraph{Global symmetries.}
Let $h^\vee$ be the dual Coxeter number of $G$.  The R-symmetry of the \Nequals1 pure super Yang-Mills theory with gauge group $G$  is $\bZ_{2h^\vee}$, such that $\bZ_2$ subgroup is the fermion number.  For simplicity, we measure the R-symmetry using $\bZ_{h^\vee}$, and take into account the fermion number separately. 
As far as we only consider pure \Nequals{1} super Yang-Mills theory, we can now set $\theta_\text{cont}=0$ by performing a phase rotation of the gaugino. 

When the gauge group is $G/K$, some of the R-symmetry is lost, since $\theta_\text{cont}\to \theta_\text{cont}+2\pi$ is no longer a symmetry.  But there is an additional symmetry that is useful.   To discuss it, let us consider for a moment a general situation where the spatial slice to be a three-dimensional manifold $X$ whose integral homology has no torsion. We later set $X=T^3$.

In the ultraviolet, note that the topological type of a $G/K$ bundle on $X$ is specified by $m\in H^2(X,K)$, which is the generalized Stiefel-Whitney class of the bundle.
By quantizing the system, the kets are labeled by $m$. On these states,
we can define  an action of $g\in H^1(X,K^\vee)$ given by \begin{equation}
g\ket{m} = \vev{g,m} \ket{m}
\end{equation}
where $\vev{\cdot,\cdot}:H^1(X,K^\vee)\times H^2(X,K)\to \U(1)$ is the natural pairing. 

In the far infrared, we just have a discrete  gauge theory with gauge group  $K^\vee$, which is partially confined to some subgroup as we will see below.  There is no matter charged under $K^\vee$.  
The dynamical variable of the system is a $K^\vee$-bundle $a$ on $X$, and we have a symmetry given by sending $a\mapsto g\otimes a$, where $g$ is another $K^\vee$-bundle.
Both $a$ and $g$  can be specified  by their holonomies $a,g\in H^1(X,K^\vee)$. In the ket notation, we have \begin{equation}
g\ket{a}=\ket{ga}
\end{equation} where we use multiplicative notation for the group structure in $H^1(X,K^\vee)$.

In the language of \cite{Kapustin:2014gua,GreatArticleToAppear}, we have a  1-form global symmetry with group $K^\vee$ in four dimensions, and it becomes an ordinary global symmetry $H^1(X,K^\vee)$ when compactified on $X$. 
When we compare the index in the ultraviolet and in the infrared, we should be able to match it including the charge $H^2(X,K)$ under this global symmetry group $H^1(X,K^\vee)$.  In fact, it is easier to do so than to count the total index itself, as we will see soon.

\paragraph{Ultraviolet.} Let us first perform  the ultraviolet computation. 
When the gauge group is $G/C$, the topological class of the gauge bundle on $T^3$ is given by an element $m=(m_{23},m_{31},m_{12})\in C^3$, where $m_{ij}$ is the generalized Stiefel-Whiteny class evaluated on the $T^2$ face in the direction $ij$.
  For each given $m$, the supersymmetric quantum mechanics on the moduli space of almost commuting triples in $G$ was performed in \cite{Witten:2000nv}, when $\theta_\text{disc}=0$.  The details of the computation depended on the choice of $G$, but in the end it was found that there are always $h^\vee$ states in total.

The resulting states carry discrete electric charges $e=(e_1,e_2,e_3)\in C^3$ 
 and the R-charge  $k\in \bZ_{h^\vee}$. But since $k\to k+ 1$ is equivalent to $e\to e+\Delta m$, we cannot measure $k$ and $e$ simultaneously.  For our purposes it is convenient to decide to measure the R-symmetry charge $k$ under the subgroup $\bZ_{h^\vee/n_m}\subset \bZ_{h^\vee}$ depending on $m$, where $n_m$ is the smallest positive integer such that $n_m\Delta m =0$. Then  we can measure the electric charges in $C^3$. Then what was found in \cite{Witten:2000nv} concerning the zero-energy states in a given sector of $m\in C^3$  can be summarized as follows, see (3.30) and (3.31) in that paper: \begin{itemize}
 \item The possible electric charge is of the form $e=k \Delta m$, \ $k=1,\ldots, n_m$.
 \item For each such $e$, and for each possible R-charge under $\bZ_{h^\vee/n_m}$, there is one state.
 \item Every zero-energy states have the same $(-1)^F=(-1)^{\rank G}$. 
\end{itemize} In total, there are indeed $h^\vee$ states in that sector.
 
With this result it is easy to count how many vacua there are when the gauge group is $G/K$ and the discrete theta angle is $\theta_\text{disc}$. The topological type of the $G/K$ bundle is classified by $m\in K^3 \subset C^3$.  The possible electric charge is then of the form $k\Delta m + \theta_\text{disc} m$ for $k=1,\ldots, n_m$.
Now, we need to impose a discrete version of the Gauss law constraint, saying that the state is uncharged under the gauge transformation in $K\subset G$. This restricts the electric charge to be zero in $G^\vee/L^\vee=K^\vee$. 
 Therefore, the number of the vacua for a given $m\in K^3$ is 
\begin{equation}
\frac{h^\vee}{n_m} \times \# \{ k=1,\ldots,n_m \mid (k\Delta+\theta_\text{disc})m=0 \in (K^\vee)^3\} \label{generalUV}
\end{equation}
which is either $h^\vee/n_m$ or $0$, depending on whether there is a $k$ such that $(k\Delta+\theta_\text{disc})m=0$ or not. In the former case, there is one state for each possible R-charge under $\bZ_{h^\vee/n_m}$.

\paragraph{Infrared.}
Next, let us perform the infrared computation.
There are $h^\vee$ vacua, with the gaugino condensate given by \begin{equation}
\vev{\tr\lambda\lambda}=\Lambda^3,\quad \omega \Lambda^3,\quad \ldots,\quad \omega^{h^\vee-1}\Lambda^3
\end{equation} where $\omega=\exp(2\pi i/h^\vee)$. They are related by the operation $\theta_\text{cont}\to\theta_\text{cont}+2\pi$.

In a theory with $G/K$ gauge symmetry, we have 't Hooft line operators whose magnetic charges are valued in $K$.    They give infrared line operators.  On a spatial slice $X$, these lines are labeled by $H_1(X,K)$.
 Let us say that the condensate is purely magnetic in the zero-th vacuum when the discrete theta angle is zero. Then all these infrared line operators have perimeter law.
 We can say that in the infrared, there is a gauge field with finite gauge group $K^\vee$, so that these line operators are electrically charged Wilson line operators of this finite group $K^\vee$.

In the $k$-th vacuum, with the discrete theta angle $\theta_\text{disc}$,
the condensate has an induced electric charge due to the operation $\theta_\text{cont}\to \theta_\text{cont}+2\pi k$ and also due to the discrete theta angle. As recalled above, the two effects can be combined, by changing the discrete theta angle by  $\theta_\text{disc} \to k\Delta + \theta_\text{disc}$.

In the infrared description we adopted above, where the line operators are electrically charged under the finite gauge field with gauge group $K^\vee$, the condensate has now the induced magnetic charge.
 The line operator with charge  $m\in H_1(X,K)$ shows the area law if it feels the induced electric charge in the condensate, i.e.~when $(k\Delta +\theta_\text{disc})m \neq 0 \in H_1(X,K^\vee)$.
This means that the gauge symmetry $K^\vee$ is partially confined to a subgroup $K^\vee_k$ which is the kernel of  $k\Delta +\theta_\text{disc}$.
We also see that it has a nontrivial confinement index in the language of \cite{Cachazo:2002zk}.

Now, let us put the infrared theory on a big $X=T^3$.
The $k$-th vacuum has a discrete $K^\vee_k$ gauge symmetry, and it gives $|K^\vee_k|^3$ zero-energy states. 
We would like to identify the charges of these states under the global symmetry $H^1(X,K^\vee)$ mentioned above.  The charge under this global symmetry is given by an element $m\in H^2(X,K)\simeq H_1(X,K)$, which can be naturally identified with the charge of the line operators. 
When $(k\Delta+\theta_\text{disc})m\neq 0$, they are confined as argued above and therefore they cost non-zero energy. When $(k\Delta+\theta_\text{disc})m=0$ they are not confined, and indeed there are $|K^\vee_k|^3$ such states.

Let us now count the number of vacua with a given $m$, varying $k$.  This is of course  \begin{equation}
\# \{ k=1,\ldots,h^\vee \mid (k\Delta+\theta_\text{disc})g=0\}\label{generalIR}
\end{equation}
which is either $0$ or ${h^\vee}/{n_m}$, where 
in the latter case,  $n_m$ is the smallest positive integer such that $n_m \Delta m=0$.
Note that the  R-symmetry  $\bZ_{h^\vee}$ in this sector is broken to $\bZ_{h^\vee/n_m}$, which rotates these states. Therefore,  there is exactly one state  for each possible R-charge under $\bZ_{h^\vee/n_g}$ in this case. 
Generalizing the argument given in footnote~\ref{IRfootnote}, we see that all the zero-energy states thus found have the same value of $(-1)^F$.

\paragraph{Comparison.} What we found in the ultraviolet \eqref{generalUV} and in the  infrared \eqref{generalIR} are clearly equal, including the charge under the R-symmetry. This is as it should be, since the Witten index is independent of the size of the box, in each of the charge sector under the global symmetry $H^1(X,K^\vee)$ of the system on $X=T^3$.

At this point we see that we did not add almost anything compared to the understanding already given in \cite{Witten:2000nv}, as already mentioned at the end of the Introduction. In the previous sections we compared the total Witten index, that looked complicated. But in fact it is easier and more trivially related to what was done in \cite{Witten:2000nv} to compare the index in a fixed value of  the charge of the  low-energy line operators   $m\in H_1(T^3,K)$ or equivalently  the ultraviolet topological class $m \in H^2(T^3,K)$.


\section*{Acknowledgements}
The author would like to thank Particle Physics Theory Group at Osaka University, for inviting him to give a series of introductory lectures on supersymmetric gauge theories; it was during the preparation of the lectures that he noticed the question addressed in this short note. 
It is also a pleasure for the author to thank O.~Aharony and N.~Seiberg for helpful comments on 1-form global symmetries, and K.~Yonekura and E.~Witten for clarifying the author's confusion on $(-1)^F$ of the states contributing to the index. 
The work of YT is  supported in part by JSPS Grant-in-Aid for Scientific Research No. 25870159,
and in part by WPI Initiative, MEXT, Japan at IPMU, the University of Tokyo.

\newpage

\bibliographystyle{ytphys}
\small\baselineskip=.9\baselineskip
\let\bbb\bibitem\def\bibitem{\itemsep1pt\bbb}
\bibliography{ref}
\end{document}